\documentclass[prl,twocolumn,superscriptaddress,showpacs]{revtex4-1}
\usepackage{graphicx,amsmath,amssymb,amsfonts}
\usepackage{textcomp}
\usepackage{array}
\usepackage{multirow}

\begin{document}

\title{A Single-Crystal Neutron Diffraction Study on Magnetic Structure of CsCo$_{2}$Se$_{2}$}

\author{Juanjuan Liu}
\author{Jieming Sheng}
\author{Wei Luo}
\author{Jinchen Wang}
\author{Bao Wei}
\email{wbao@ruc.edu.cn}
\affiliation{Department of Physics, Renmin University of China, Beijing
	100872, China }
\author{Jinhu Yang}
\affiliation{Department of Physics, Hangzhou Normal University, Hangzhou 310036, China}
\author{Minghu Fang}
\affiliation{Department of physics Zhejiang University, Hangzhou 310027, China}
\affiliation{Collaborative Innovation Center of Advanced Microstructures, Nanjing 210093, China}
\author{S.A. Danilkin}
\affiliation{Bragg Institute, ANSTO, Locked Bag 2001, Kirrawee DC NSW 2232, Australia}

\date{\today}

\begin{abstract}

The magnetic structure of CsCo$_{2}$Se$_{2}$ was investigated using single-crystal neutron diffraction technique. An antiferromagnetic transition with the propagation vector
(0,0,1) was observed at T$_{N} = 78$ K. The Co magnetic moment 0.772(6) $\mu_B$ at 10 K pointing in the basal plane couples ferromagnetically in the plane which stacks antiferromagnetically along the $ c $ direction. Tuning and suppressing the interplane
antiferromagnetic interaction may be crucial to induce  a superconducting state in the material.

\end{abstract}

\maketitle

\begin{figure}[!htbp]
     \includegraphics [width=0.6\columnwidth]{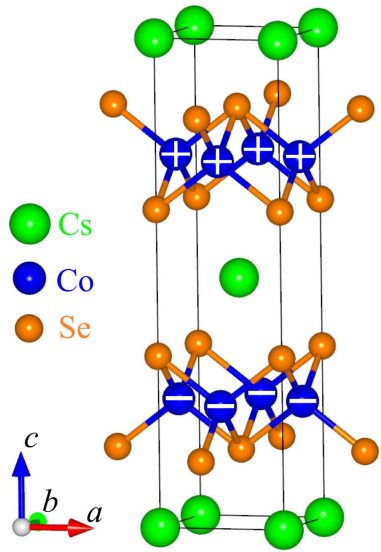}
     \caption{(color online)  The crystal and magnetic structure of CsCo$_{2}$Se$_{2}$ in the $I4/mmm$ unit cell. The magnetic moments of the cobalt atoms are pointed in the $ab$ plane, but the moment direction within the plane can not be distinguished in a diffraction study due to the tetragonal symmetry of the crystal. The phase of antiferromagnetic moments, therefore, is denoted by + and - signs.}
     \label{fig:str}
\end{figure}

The discovery of superconductivity at quite high a transition temperature T$_c=38$~K \cite{PhysRevLett.101.107006} in the iron-based laminar pnictide compound BaFe$_{2}$As$_{2}$ doped from an antiferromagnetic parent state \cite{A062776,A071525,A071077,A073950,A074096} has raised a lot of interest in ternary chalcogenide compounds $AM_{2}X_{2}$ ($A$ = K, Rb, Cs, Tl; $M$ = Fe, Co, Ni; $X$ = S, Se) of the same ThCr$_{2}$Si$_{2}$ crystalline structure with the $I4/mmm$ space group symmetry, see Fig.~\ref{fig:str}. 
The iron chalcogenides $A_{x}$Fe$_{2-y}$Se$_{2}$ have indeed been discovered as superconductor at T$_c\approx 30$~K \cite{PhysRevB.82.180520,Maziopa.2011,Yoshikazu.2011,mhfang.2011,C125525,hdwang.2011}, however, the Fe layer contains ordered vacancy and the correct 
crystal structure for this family of Fe-based superconductors is of the $I4/m$ space group symmetry instead \cite{D014882,D020488,D023674}. Superconductivity coexits with a large-moment block antiferromagnetism in all these $A_{x}$Fe$_{2-y}$Se$_{2}$ superconductors \cite{D020830,D022882}.
While Ni chalcogenides generally show enhanced heavy fermion-like Pauli paramagnetism and become superconductor in KNi$_{2}$Se$_{2}$ (T$_c = 0.8$~K), KNi$_{2}$S$_{2}$ (T$_c = 0.46$~K) and TlNi$_{2}$Se$_{2}$ (T$_c = 3.7$~K) \cite{PhysRevB.86.054512,PhysRevB.87.045124,wang.2013,feng.2012,mhfang.2016}, 
all Co chalcogenides investigated so far remain metallic conductor. 

Most Co chalcogenides order ferromagnetically with the easy axis of the Co magnetic moments in the $ab$ plane except 
CsCo$_{2}$Se$_{2}$ and TlCo$_{2}$Se$_{2}$ which order antiferromagnetically \cite{Huan-CsCo2Se2}. 
TlCo$_{2}$Se$_{2}$ has been shown to have an incommensurate helical structure, in which the Co moments order ferromagnetically within the $ab$ plane and rotate about 121\textdegree\, between adjacent Co layers \cite{PhysRevB.70.024407,Jeong2007}. 
The interlayer magnetic interaction can be tuned by chemical tuning.
Substituting Co with Ni, adding electrons, will gradually convert the IC structure to a collinear antiferromagnetic structure \cite{Newmark-TlCoNiSe2} and then to paramagnetism at the end composition TlNi$_{2}$Se$_{2}$ which is a superconductor as stated above \cite{wang.2013}. Substituting Co with small amounts of Cu or Fe will not change the
IC magnetic structure in TlCo$_{2-x}$(Cu,Fe)$_{x}$Se$_{2}$ ($ x \leq $ 0.2) \cite{Ronneteg2006204}.

Substituting Tl by K in TlCo$_{2}$Se$_{2}$, changing the interlayer space, will similarly convert the IC magnetic structure to a collinear antiferromagnetic order, but the end compound KCo$_{2}$Se$_{2}$ orders ferromagnetically \cite{Huan-TlKCo2Se2} instead of becoming a superconductor.
Substitution at the Se site by S, which changes the crystal field of Co, the IC structure continueously changes to the ferromagnetic order by the gradual decrease of the turning angle between the adjacent Co layers \cite{Ronneteg200653, Ronneteg2010681}.
It seems that tuning the IC to commensurate antiferromagnetic order, then to a paramagnetic metal is the crucial route to superconductivity. 

The other antiferromagnetic member of the Co chalcogenides, 
CsCo$_2$Se$_2$, has been investigated previously only with powder neutron diffraction \cite{Fabian-CsCo2Se2}.
Since the Cs containing sample is very air sensitive, single crystalline sample is advantageous over ground powders.
In this work, we conducted neutron diffraction investigation on magnetic structure and transition using a single-crystal CsCo$_2$Se$_2$ sample. 
The antiferromagnetic transition occurs at $T_N= 78$ K which is consistent with bulk property measurements \cite{PhysRevB.88.064406}.
The antiferromagnetic structure consists of ferromagnetically ordered Co moments in the  Co$_2$Se$_2$ layer with the magnetic moments orient in the $ab$ plane, and the Co layers are coupled antiferromagnetically as shown in Fig.~\ref{fig:str}.
Magnetic moment $ M = $ 0.772(6) $ \mu_{B} $ per Co at 10 K,
which is substantially larger than that from the previous neutron powder diffraction work \cite{Fabian-CsCo2Se2}.

The single crystal sample of CsCo$_{2}$Se$_{2}$ was grown by the self-flux method \cite{PhysRevB.88.064406} with dimensions $\sim 5 \times 3 \times 0.3$ mm$^{3}$. 
The grown sample was sealed in an evacuated quartz tube for it is very sensitive to moisture and air.
The single crystal still showed metallic luster during our transfer in a glove box of the sample to a helium sealed sample can for neutron diffraction experiments. 
Neutron diffraction measurements were performed using the thermal triple-axis spectrometer Taipan at ANSTO \cite{Danilkin-triple-axis}. 
The horizontal collimations were open-40$^{'}$-40$^{'}$-open. 
Neutrons with incident energy $E = 14.7$ meV were selected using the (0, 0, 2) reflection of a pyrolytic graphite (PG) monochromator. 
PG filters were used to remove higher-order neutrons from the neutron beam. The temperature of the sample was regulated by the ILL Orange Cryostat in a temperature range from 10 to 300 K. 

CsCo$_{2}$Se$_{2}$ crystallizes in the ThCr$_{2}$Si$_{2}$ body-centered tetragonal structure (space group No. 139, $I4/mmm$) \cite{Huan-CsCo2Se2}. The lattice parameters
of our sample were determined to be $a = 3.914$ {\AA} and $c = 14.85$ {\AA} at 10 K.
As shown in Fig.~\ref{fig:str}, each magnetic atom Co is four-fold coordinated by As atom either above or below the layer to form a tetrahedron.
The edge sharing tetrahedrons form the Co$ _{2} $Se$ _{2} $ layers which are separated by the non-magnetic cesium cations. 
The single crystal sample was aligned in the $(hhl)$ scattering plane during this neutron diffraction study. 
In the $(hhl)$ scattering plane, nuclear Bragg peaks were observed at the even $l$ positions, which is consistent with the selection rule of the $I4/mmm$ space group, $ h + k + l  = 2n $. 
Below 78 K, a new set of magnetic peaks show up at odd $l$ positions, which are shifted from the nuclear Bragg peaks by the magnetic propagation vector $\textbf{Q}_{AF} = (0, 0, 1)$. 
This situation facilitates the collection and measurements of the separated Bragg peaks of crystalline and magnetic origins.

\begin{figure}[!htbp]
	\centering
     \includegraphics [width=0.9\columnwidth]{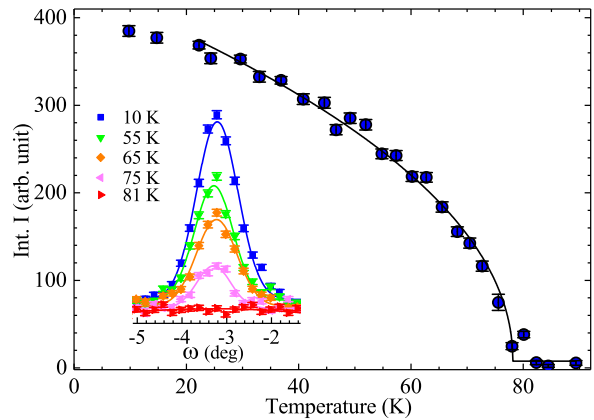}
     \caption{(color online) Temperature dependence of the (0,0,1) magnetic Bragg peaks. The solid circles are experimental data with error bar. And the solid line is a fit of the experimental data by the function $I = I_{0} + A(1-T/T_{N})^{2\beta}$. Inset: The rocking scans of the (0,0,1) Bragg reflection measured at 10, 55, 65, 75 and 81K, with increasing temperature.}
     \label{fig2}
\end{figure}

To investigate the magnetic phase transitions, we measured the temperature dependence of magnetic Bragg reflection (0, 0, 1) from base temperature 10 K to 87 K, refer to Fig.~\ref{fig2}. 
It clearly illustrates the emergence of the antiferromagnetic phase below the transition temperature. 
The measured intensity was fitted to the function $ I = I_{0} + A(1-T/T_{N})^{2\beta}$ to obtain the critical temperature $T_{N} = 78.1(3)$ K. 
It is consistent with magnetic susceptibility measurements of the CsCo$_{2}$Se$_{2}$ single crystal samples \cite{PhysRevB.88.064406, Fabian-CsCo2Se2}, but it is 10 K higher than that reported in previous powder neutron diffraction work \cite{Fabian-CsCo2Se2}. 
The rocking scan in the inset of Fig.~\ref{fig2} has the resolution-limited peak width and shows no detectable change with temperature. 
It indicates the long-range nature of the antiferromagnetic ordering.

The propagation vector $\textbf{Q}_{AF} = (0, 0, 1)$ implies an antiferromagnetic spin arrangement. 
The little group $ G_{k} $ for (0, 0, 1) under the space group of $I 4/mmm $ is the same as the crystal point group 4$ /mmm $. 
The magnetic representation for $ G_{k} $ could be decomposed into 10 irreducible representations (IRs) including eight one-dimensional IRs $\Gamma_{1} \dots \Gamma_{8}$ and two two-dimensional IRs $\Gamma_{9}$ and  $\Gamma_{10}$  \cite{ritter2011}.
For Co atom on the crystallographic 4$ d $ (0, 0.5, 0.25) position, it gives four symmetry allowed solutions as follow: 
\begin{equation}\label{eq:IRs}
\Gamma_{Mag} = 1 \Gamma_{2} + 1 \Gamma_{5} + 1 \Gamma_{9} + 1 \Gamma_{10}.
\end{equation}

\begin{table*}[!htbp]
	\caption{(color online) The basis vectors (BV) of decomposed irreducible representations (IR) of space group $I 4/mmm $ (No. 139) with wave-vector $(0, 0, 1)$ and moments on $4d$ site. The corresponding magnetic structures are displayed.}
	\label{tab:irrep}
	\begin{tabular}{m{2cm}<{\centering}*{4}{m{3.8cm}<{\centering}}}
		\hline\hline
		\multirow{2}{*}{Irrep}  & \multicolumn{4}{c}{Co at $ 4d $ site, $\boldsymbol{k} = (0, 0, 1)$} \\
		\cline{2-5}  & $ \Gamma_{2} $ & $ \Gamma_{5} $ & $ \Gamma_{9} $ & $ \Gamma_{10} $  \\
		\hline BV for site \  (0,0.5,0.25) &  $(0, 0, m_z)$  &  $(0, 0, m_z)$  &  $(m_x, 0, 0)$  $(0, m_y, 0)$  &  $(m_x, 0, 0)$  $(0, m_y, 0)$    \\
		BV for site \  (0,0.5,0.75) & $(0, 0, -m_z)$  &  $(0, 0, m_z)$  &  $(-m_x, 0, 0)$  $(0, -m_y, 0)$  &   $(m_x, 0, 0)$  $(0, m_y, 0)$    \\
		\hline magnetic structure mode  &  {\includegraphics[width=15mm]{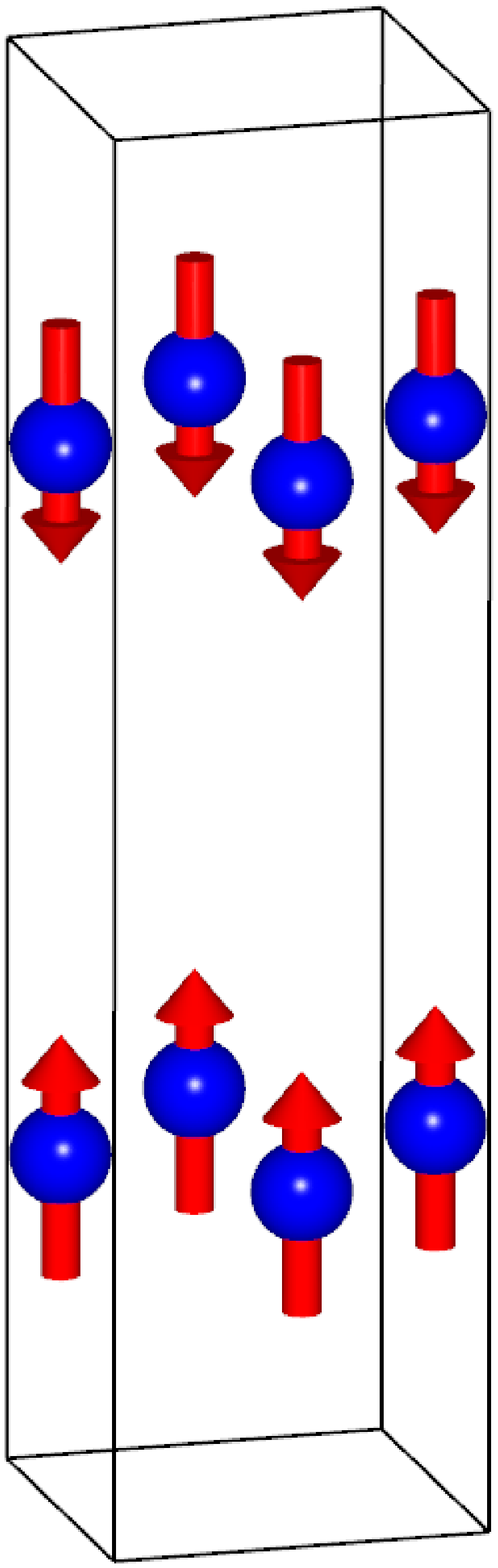}} &  {\includegraphics[width=15mm]{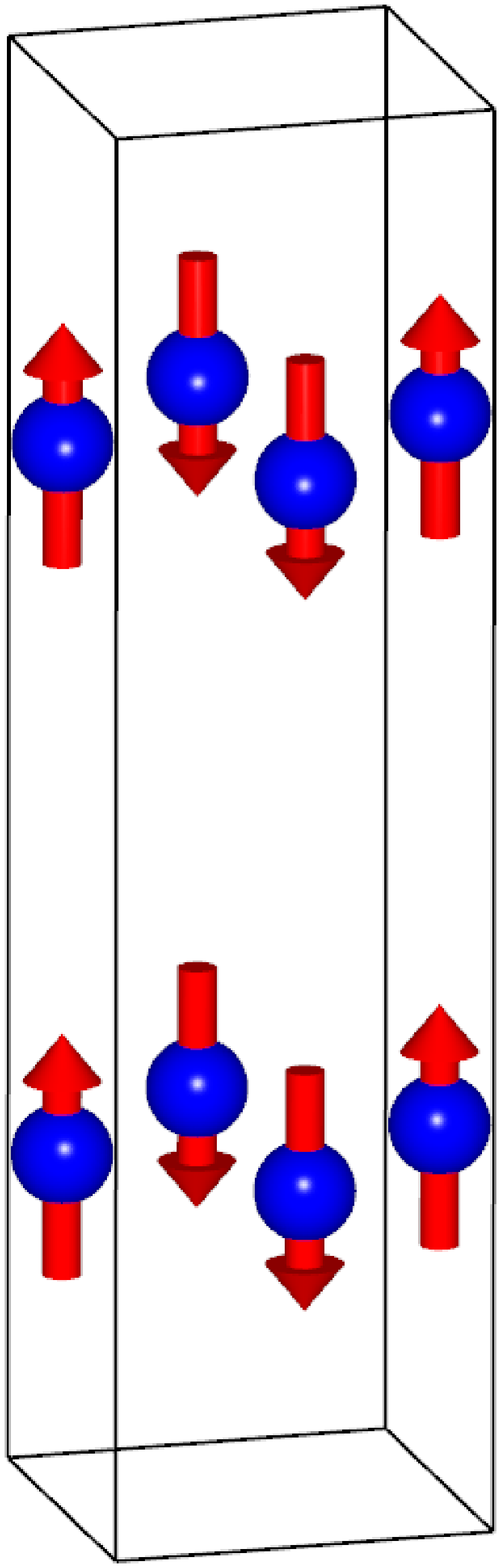}} &  {\includegraphics[width=15mm]{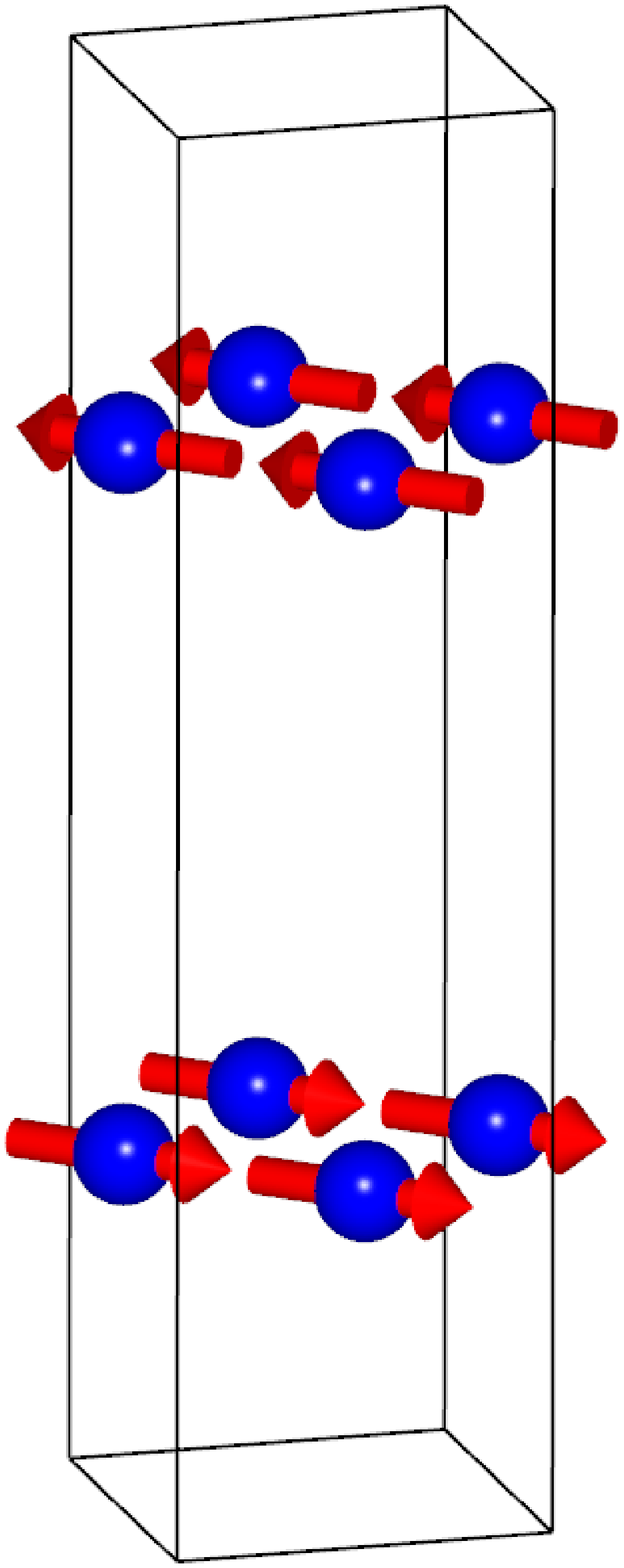}} &  {\includegraphics[width=15mm]{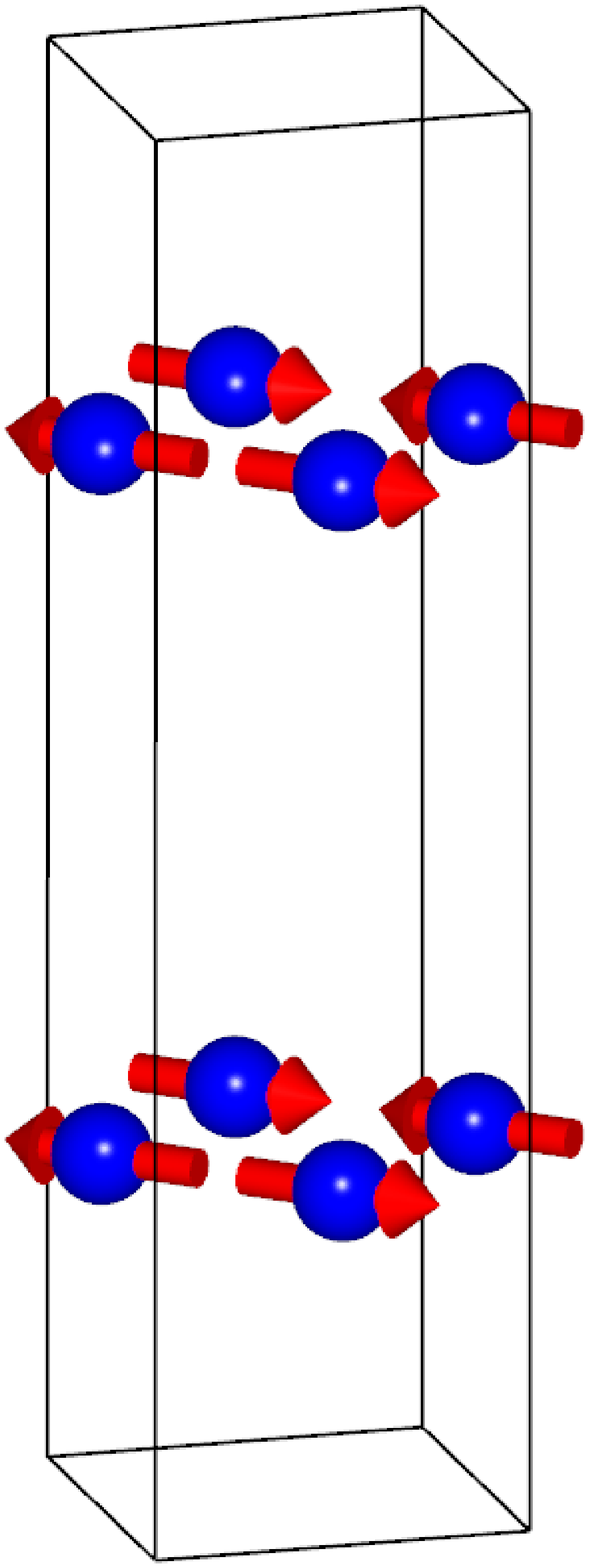}}  \\ 
		\hline\hline
	\end{tabular}
\end{table*}

The basic vectors and  spin configurations corresponding to the four IRs are shown in Table \ref{tab:irrep}. 
The one-dimensional IRs $ \Gamma_{2} $ and $ \Gamma_{5} $ give the magnetic structure with magnetic moments aligned along the crystallographic $ c $ axis. 
For the magnetic structure models given by the other two two-dimensional IRs, the moments are constrained in the $ ab $ plane. 
Among the four magnetic configurations, $ \Gamma_{5} $ and $ \Gamma_{10} $ have a chess board like spin configuration in the $ ab $ plane and arrange ferromagnetically along the $ c $ direction, which give a zero magnetic structure factor on the $ l = $ odd reflections, inconsistent with our observations.
Whereas $ \Gamma_{2} $ and $ \Gamma_{9} $ give a ferromagnetic spin arrangement in the $ ab $ plane but align antiferromagnetically along the  $ c $ direction, for which the magnetic structure factor of the $ (h, h, l) $ reflection is a non-zero constant, 16. 
Hence, $ \Gamma_{2} $ and $ \Gamma_{9} $ represent the possible magnetic structure which is commonly referred to as the A-type antiferromagnetic structure.

To distinguish the $ \Gamma_{2} $ and $ \Gamma_{9} $ magnetic structures in CsCo$_{2}$Se$_{2}$, we need to consider the details of the various magnetic Bragg peaks.
For a commensurate antiferromagnetic structure as indicated by the $\textbf{Q}_{AF}$,
the magnetic neutron scattering cross section of CsCo$_{2}$Se$_{2}$ is described by 
\begin{equation}\label{eq:cross_section}
\sigma (\textbf{q}) = (\frac{\gamma r_{0}}{2})^{2} N_{m} \dfrac{(2 \pi)^{3}}{v_{m}} \langle M \rangle ^{2} \vert f(q)\vert ^{2} \vert \mathcal{F}\vert ^{2} \langle 1 - (\hat{\textbf{q}} \cdot \hat{\textbf{s}})^{2} \rangle ,
\end{equation}
where $(\gamma r_{0}/2)^{2} = $ 0.07265 barn/$\mu_{B}^{2} $, $N_{m}$ is the number of magnetic unit cells in the sample, $v_{m}$ the volume of magnetic unit cell, $M$ the magnetic moment of Co, $f(\textbf{q})$  the magnetic form factor of the Co ion, $ \hat{\textbf{q}} $ the unit vector of $\textbf{q}$, $ \hat{\textbf{s}} $ the unit vector of the Co moment, and $\mathcal{F}$ the magnetic structure factor per unit cell  \cite{squires}. $N_{m}$ and $v_{m}$ will be normalized out by measurements of structural Bragg peaks.

The polarization factor $ \langle 1 - (\hat{\textbf{q}} \cdot \hat{\textbf{s}})^{2} \rangle $ is averaged over magnetic domains.
For an arbitrary moment orientation, there are in general 16 magnetic $ M $ domains with the tetragonal symmetry of CsCo$_{2}$Se$_{2}$.
The moment direction within the basal plane cannot be distinguished by diffraction due to the square symmetry \cite{Shirane:a02507}.
Denote $ ( h, h, l) $ as the Miller indices of the neutron scattering vector $ \textbf{q} = (h\frac{2\pi}{a}, h\frac{2\pi}{a}, l\frac{2\pi}{c})$, the angle between magnetic moment and the $ c $ axis as $ \beta $, then we have 
\begin{equation}\label{eq:qs_factor_general}
\langle1-(\hat{\textbf{q}}.\hat{\textbf{s}})^{2}\rangle=1-\frac{\left(\frac{h}{a}\right)^2 \sin^2\beta + \left(\frac{l}{c}\right)^2 \cos^2\beta}{2\left(\frac{h}{a}\right)^2+\left(\frac{l}{c}\right)^2}.
\end{equation}

For a magnetic moment along the $ c $ axis corresponding to the model of IR $ \Gamma_{2} $, $ \beta = $ 0\textdegree, the polarization factor is reduced to 
\begin{equation}\label{eq:qs_factor2}
\langle1-(\hat{\textbf{q}}.\hat{\textbf{s}})^{2}\rangle=1-\frac{\left(\frac{h}{a}\right)^2}{2\left(\frac{h}{a}\right)^2+\left(\frac{l}{c}\right)^2}.
\end{equation}
For that of IR $ \Gamma_{9} $, $ \beta = $ 90\textdegree, the polarization factor is reduced to 
\begin{equation}\label{eq:qs_factor}
\langle1-(\hat{\textbf{q}}.\hat{\textbf{s}})^{2}\rangle=1-\frac{\left(\frac{l}{c}\right)^2}{2\left(\frac{h}{a}\right)^2+\left(\frac{l}{c}\right)^2}.
\end{equation}

Comparing the calculated magnetic cross sections of the magnetic configurations represented by $ \Gamma_{2} $ and $ \Gamma_{9}$ with our measured magnetic cross sections yielded from the integrated intensity of rocking scans by $\sigma(\textbf{q})=I(\textbf{q})\sin(2\theta)$ and normalized by a series of structural Bragg peaks,
the magnetic structure corresponding to $ \Gamma_{9}$ with moment lying in the $ ab $ plane is found to be the chosen one.
The measured magnetic cross section is presented in Table \ref{tab:mag_cross_section} in good agreement with the calculated magnetic cross section from the magnetic structure represented by IR $ \Gamma_{9}$ with the reliability coefficient $ R $ being 8.3\%.
The magnetic moment is determined at 10 K to be $ M = $ 0.772(6) $ \mu_{B} $ per Co atom. 
The much reduced moment from the atomic limit value of the Co ion indicates a strong itinerant magnetic behavior in CsCo$ _{2} $Se$ _{2} $. Like T$_N$, our value for $M$ is 
also larger than the 0.2$ \mu_{B} $/Co reported previously  \cite{Fabian-CsCo2Se2}.
A likely reason for the discrepancy may be attributed to the moisture and air affected powder sample used in the previous work.

\begin{table}[!tbhp]
	\caption{Magnetic Bragg intensity, $\sigma_{obv}$, defined in Eq. (\ref{eq:cross_section}), observed at 10 K, in units of barn per unit cell. The theoretical intensity, $\sigma_{cal}$, is calculated using Eqs. (\ref{eq:cross_section}) and (\ref{eq:qs_factor}) with $ M = 0.772 \mu_{B} $/Co. The resultant reliability coefficient $R = \sum\|\sigma_{obv}-\sigma_{cal}\|/\sigma_{obv}$ is $8.3\%$ }
	\label{tab:mag_cross_section}
	\centering
		\begin{ruledtabular}  
			\begin{tabular}{ccc}
  \textbf{q}        &   $\sigma_{obs}$  &   $\sigma_{cal}$\\
		\hline
		(0 0 1)    &      62(2)       &       60.3\\
		(0 0 3)    &      50(1)       &       51.4\\
		(0 0 5)    &      35(4)       &       37.8\\
		(0 0 7)    &      21(2)       &       24.4\\
		(0 0 9)	   &      8(1)         &       14.2\\
		(1 1 1)    &      27(6)       &       17.8\\
		(1 1 3)    &      28(6)       &       18.4\\
		(1 1 7)    &      10(2)       &       12.2\\
		(1 1 -1)   &      27(3)       &       17.8\\		
		(-1 -1 -1) &      17(1)       &       17.8\\
		(-1 -1 1)  &      19(1)       &       17.8\\
		(1 1 -3)   &      29(3)       &       18.4\\
		(-1 -1 3)  &      21(1)       &       18.4\\
			\end{tabular}
	\end{ruledtabular}
\end{table}

The ThCr$ _{2} $Si$ _{2} $ structure contains two radically different magnetic interaction distances. 
For $A$Co$ _{2} $$X _{2} $ ($A$ = K, Rb, Cs, Tl; $X$ = S, Se), the intralayer Co-Co distance (2.705 - 2.768 {\AA}) is comparable to that in Co metal (2.506 {\AA}) \cite{Taylor:a00276} which is appropriate for the direct interaction to form a ferromagnetic Co layer, 
while the interlayer Co-Co distance (6.797 - 7.423 {\AA}) is too large for any direct interaction.
In consequence, the intraplane ferromagnetic Co-Co correlation is quite robust while the weaker interplane magnetic interaction is subjected to chemical tuning,
in consistent with the extensive results from studies on the ternary transition metal chalcogenides \cite{Huan-CsCo2Se2, Newmark-TlCoNiSe2, Huan-TlCoNiS2, Huan-TlKCo2Se2, Mark-TlCo2SSe, Broddefalk20001317, Fabian-CsCo2Se2,Guo-KCo2SeS}. 
The RKKY (Ruderman-Kittel-Kasuya-Yoshida) exchange coupling which is very sensitive to the interlayer distance and electron density is suggested to explain the interplane interaction mechanism since all the $A$Co$ _{2} $$X _{2} $ members are metallic.
The oscillatory character of the RKKY exchange constant makes the ferromagnetic and antiferromagnetic interlayer interactions both possible.

In addition, the local tetrahedron shape, which is known to affect the T$_c$ in Fe-based superconductors \cite{A062195,A063821}, is thought to be also important in the determination
of the magnetic structure.
It's quite interesting to note that in Tl$_{1-x}$K$_{x}$Co$_{2}$Se$_{2}$ the AFM to FM transition occurs when the Co-Se-Co angle approaches the ideal angle 109.4\textdegree\, \cite{Huan-TlKCo2Se2}.
In TlCo$_{2}$Se$_{2-x}$S$_{x}$, with the increase of S content, the CoSe$_{4}$ tetrahedron is compressed gradually, leading to a reduced turning angle of helix magnetic structure from 121\textdegree\, to 0\textdegree\, \cite{Ronneteg200653}.
It is also worth to note that the CoSe$_{4}$ tetrahedron in antiferromagnetic CsCo$_{2}$Se$_{2}$ \cite{Huan-CsCo2Se2} and TlCo$_{2}$Se$_{2}$ \cite{Huan-TlKCo2Se2} is strongly elongated, while that of the ferromagnetic CsCo$_{2}$S$_{2}$ \cite{Huan-CsCo2Se2} and TlCo$_{2}$S$_{2}$ \cite{Huan-TlCoNiS2}  are compressed.
Based on these knowledge, future studies on Co chalcogenides under high pressure would be illuminating to reveal the magnetic interaction between the Co layers.

In conclusion, we investigated magnetic transition and magnetic structure of CsCo$_{2}$Se$_{2}$ in this single-crystal neutron diffraction study. 
The A-type collinear antiferromagnetic order is consistent with that reported previously in a powder neutron diffraction study. The Co magnetic moments are aligned ferromagnetically in the plane with the easy axis also in the plane. From plane to plane, the Co moments alternate in their direction. Our measured $ T_{N} = $ 78.1(3) K and the
magnetic moment 0.772(6)~$\mu_B$ per Co are both substantially larger than those reported previously in the powder neutron diffraction work \cite{Fabian-CsCo2Se2}, possibly due to
the air and moisture affected powder sample. Suppressing the collinear antiferromagnetic order may be the route to explore the possibility of the superconductivity.

The work at RUC was supported by the National Basic Research Program of China (Grant No.~2012CB921700), the National Natural Science Foundation of China (Grant No.~11190024), the Fundamental Research Funds for the Central Universities and the Research Funds of Renmin University of China grant (Grant No.~17XNLF04 and 17XNLF06). J.S.~acknowledges support from China Scholarship Council.

\end{document}